\documentclass[traditabstract]{aa}
\usepackage{natbib,amstext,textcomp,graphicx,txfonts}
\usepackage{txfonts}
\bibpunct{(}{)}{;}{a}{}{,}

\newcommand{\mum}{\,\hbox{\textmu{}m}}
\newcommand{\cm}{\,\hbox{cm}}
\newcommand{\mm}{\,\hbox{mm}}
\newcommand{\nm}{\,\hbox{nm}}

\newcommand{\au}{\,\hbox{au}}

\newcommand{\K}{\,\hbox{K}}

\begin{document}

\title{Absorption of crystalline water ice in the far infrared at different temperatures}
\author{C.~Reinert \and H.~Mutschke \and A.~V.~Krivov \and T.~L{\"o}hne \and P.~Mohr}
\institute{Astrophysikalisches Institut und Universit{\"a}tssternwarte,
           Friedrich-Schiller-Universit{\"a}t Jena,
           Schillerg{\"a}{\ss}chen 2--3, 07745 Jena, Germany, \email{caroline.reinert@uni-jena.de}
          }

\abstract{
The optical properties of ice in the far infrared are important for models of protoplanetary and debris disks. In this report, we derive a new set of data for the absorption (represented by the imaginary part of the refractive index $\kappa$) of crystalline water ice in this spectral range. The study includes a detailed inspection of the temperature dependence, which has not been conducted in such detail before. We measured the transmission of three ice layers with different thicknesses at temperatures $\vartheta = 10\ldots250\K$ and present data at wavelengths $\lambda=80\ldots625\mum$. We found a change in the spectral dependence of $\kappa$ at a wavelength of $175 \pm 6 \mum$. At shorter wavelengths, $\kappa$ exhibits a constant flat slope and no significant temperature dependence. Long-ward of that wavelength, the slope gets steeper and has a clear, approximately linear temperature dependence. This change in behaviour is probably caused by a characteristic absorption band of water ice.
The measured data were fitted by a power-law model that analytically describes the absorption behaviour at an arbitrary temperature. This model can readily be applied to any object of interest, for instance a protoplanetary or debris disk. To illustrate how the model works, we simulated the spectral energy distribution (SED) of the resolved, large debris disk around the nearby solar-type star HD\,207129. Replacing our ice model by another, commonly used data set for water ice results in a different SED slope at longer wavelengths. This leads to changes in the characteristic model parameters of the disk, such as the inferred particle size distribution, and affects the interpretation of the underlying collisional physics of the disk.
}
\maketitle
%
\section{Introduction} \label{sec:intro}
Water in its different forms is ubiquitous in star- and planet-forming regions \citep[and references therein]{vandishoeck2014} as well as in actual planetary systems like the solar system \citep{encrenaz2008}. Water vapour was observed in 24\% of gas-rich disks around T Tauri stars \citep{riviere2012} and in protoplanetary disks such as TW Hydrae \citep{hogerheijde2011}. Water ice can be observed in all these disks \citep[e.\,g.][]{waters1996}. On the other hand, liquid water is only assumed to exist in certain objects within the solar system \citep[e.\,g.][]{tobie2008,kalousova2014}, but is generally hard to detect remotely. Especially when a disk gets older, i.\,e. it transforms into a debris disk, we expect only solid state forms to survive the sublimation and dissociation caused by stellar radiation.

There are two basic types of water ice: crystalline and amorphous. Which of the two ice types occurs depends on the temperature at which it has formed. Ice formed at temperatures below $110\K$ acquires a totally amorphous structure, while at higher temperatures it becomes crystalline \citep{smith1994,moore+hudson1992}. If amorphous ice is warmed above $110\K$, it will also become crystalline. But the transformation occurs not at this single temperature, but over a range up to $130\K$. \citep[e.\,g.][]{schmitt1989}.
Crystallization is an irreversible process, which leads to the suggestion that water ice in a crystalline form dominates even in cold regions of protolanetary systems. Because of radial mixing, the crystalline water ice, which has possibly formed in the warmer regions of the disk near the star, could have been spread over the whole disk. A similar scenario is used to explain the existence of crystalline silicates in colder regions of protoplanetary disks \citep[e.\,g.][]{boekel2004}. On any account, crystalline water ice was found in several protoplanetary disks around young stars \citep[e.\,g.][]{terada+tokunaga2012,mcclure2012}.

As successors of the described protoplanetary disks, debris disks must contain crystalline water ice as well because the dust in them is produced by collisions between small bodies, which have initially formed in the protoplanetary stage of the system. Some crystalline minerals have already been found, notably crystalline silicates in the environment of $\beta$~Pictoris \citep[e.g.][]{vries2012}. Yet no clear observational evidence for water ice in debris disks has been found so far. However, including crystalline water ice into the assumed composition for debris dust has been shown to improve models of debris disks in many cases, resulting in a better fit to the observations \citep[e.\,g.][]{chen2008,reidemeister2011,lebreton2012,donaldson2013}.

Therefore, the optical properties of ice are important for understanding of the physics of planetary and protoplanetary systems. The far-infrared spectral range ($\lambda > 50\mum$) is especially interesting since the thermal radiation emitted by such cold objects peaks at these wavelengths. There have been many experimental campaigns to measure optical data (for absorption in particular) of water ice in the infrared. Most of these campaigns cover the near-infrared range \citep[e.\,g.][]{grenfell+perovich1981,kou1993} and the region of the strong absorption bands of water ice around $44$ and $63\mum$ \citep{moore+hudson1992,smith1994,johnson+atreya1996,schmitt1998}. The shapes of the features at $44$ and $63\mum$ has been found to differ for amorphous and crystalline ice. In the latter, two sharp peaks at both positions are clearly seen, whereas in amorphous ice there is one smaller and broader peak at around $44\mum$ with only a shoulder at the longer wavelength end. While the feature region is well explored, there were only a few investigations of wavelengths beyond them ($\lambda \geq 100\mum$).

In this paper, we  measure the absorption of crystalline water ice at $45\mum < \lambda < 1000\mum$ for six different temperatures between $10\K$ and $250\K$. We approximate the behaviour by power-law models and take particular care of the temperature dependence of the absorption index $\kappa$ and a previously noted water ice feature at about $166\mum$ generated by intermolecular bending modes \citep{wozniak+dera2007}, because this feature is not yet discussed in the astronomical literature.

Section~\ref{sec:previous_work} gives a short overview of the previous measurements on ice in the far infrared. The measurement techniques and the treatment of the data are described in section~\ref{sec:methods}. The results are presented in section~\ref{sec:results}. In section~\ref{sec:discussion}, we compare our results to previous data sets and discuss possible future measurements. To illustrate how our results can be applied to specific objects, in section~\ref{sec:application} we use them to simulate the debris disk of HD\,207129. Section~\ref{sec:conclusions} lists our conclusions.
%
\section{Previous work} \label{sec:previous_work}
Investigations on the absorption of water ice for wavelengths beyond $100\mum$ are rare and differ substantially in quality. Some of these data sets comprise very few data points \citep[see][]{irvine+pollack1968,bertie1969}, some spread over more than an order of a magnitude \citep[see][]{hudgins1993}, and some only give relative values \citep[see][]{bertie+whalley1967}. Accordingly, for further inspection we only selected three data sets free of these shortcomings:
\begin{itemize}
\item
\citet{curtis2005} condensed water vapour on a cooled substrate to produce an ice layer. They measured at the temperature at which the ice had formed. These temperatures ranged from $106\K$ to $176\K$ where between the measurements at $126\K$ and $136\K$ the ice transforms from amorphous to crystalline state. Additionally, they adjusted the layer thickness to the wavelength region of the measurement. For the far-infrared region the thickest layers were used to obtain an appropriate absorption signal. Altogether, their data cover the range from $15\mum$ to $192\mum$.
\item
\citet{whalley+labbe1969} froze distilled water in their experiments and measured in a wavelength region of $\lambda=235\ldots455\mum$. This technique always resulted in polycrystalline water ice, independent from the temperature.
\item
\citet{mishima1983} performed their measurements with monocrystalline ice cubes at wavelengths from $400\mum$ to $1.25\mm$. The technique used to produce monocrystalline ice is not described in their study.
\end{itemize}
\begin{figure}[t!]
  \centering
  \includegraphics[width=0.7\linewidth,angle=270]{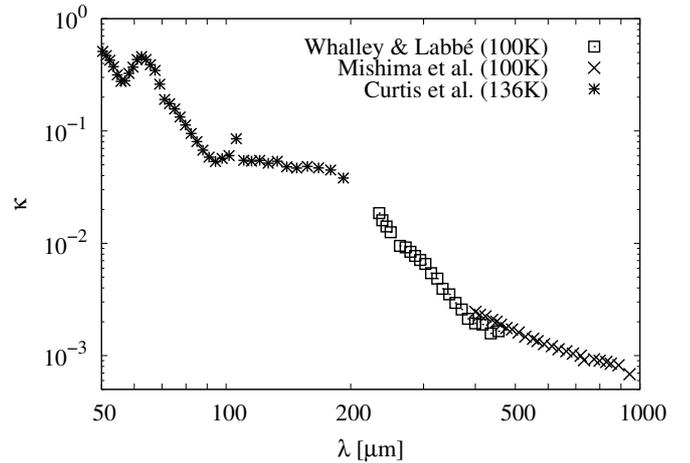}\\%
  \caption{Data for the imaginary part $\kappa$ of the refractive index $N=n+i\kappa$. Shown are data from \citet{curtis2005}, \citet{mishima1983} and \citet{whalley+labbe1969} for crystalline water ice near $100\K$.
  \label{fig:data_literature}}
\end{figure}

These three data sets are visualized in Fig.~\ref{fig:data_literature} for temperatures near $100\K$ and crystalline ice only. The figure plots the imaginary part $\kappa$ of the refractive index $N=n+i\kappa$ depending on the wavelength. We have chosen $\kappa$ because it is a material parameter independent of the geometry of the sample.

\begin{figure}[t!]
	\centering
	\includegraphics[width=0.7\linewidth, angle=270]{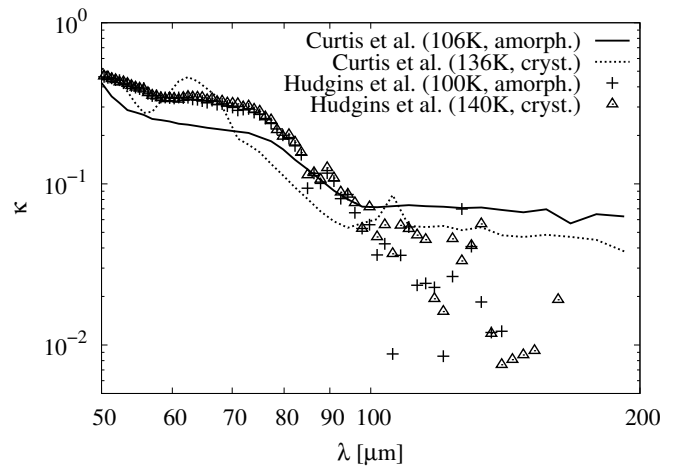}\\%
	\caption{Imaginary part $\kappa$ of the refractive index $N$ for different temperatures and ice types from \citet{curtis2005} and \citet{hudgins1993}.
	\label{fig:vgl_curtis}}
\end{figure}

Most of these data are for crystalline ice. Only \citet{hudgins1993} and \citet{curtis2005} measured both ice types beyond the feature area. The Curtis et al. data reveal no significant difference between the two ice types, except for the features (Fig.~\ref{fig:vgl_curtis}).
The data of Hudgins et al. show a huge scatter for $\lambda>100\mum$, but also no particular difference between the amorphous and crystalline ice type. In this case, there is not even a difference in the feature region to be seen. This might lead to the suggestion that Hudgins et al. measured amorphous ice only, despite the temperature of $140\K$. For $\lambda > 200\mum$, no measurements of both ice types have been performed. As a result, a comprehensive comparison of amorphous and crystalline ice is hard to make.

To overcome the problems of the measurements and to alleviate applications of the laboratory data to astronomical objects, there have been some attempts to create an artificial model to describe the optical properties of ice.

\citet{warren1984} combined multiple data  sets for crystalline water ice into one, ranging from the ultraviolet to the radio wave region ($44.3\nm$ to 8.6\,m). To obtain a data set for a temperature of $\vartheta=213\K$~$(-60^\circ C)$ he used the difference in absorption between the $100\K$ and $200\K$ measurements from the data of \citet{mishima1983} and \citet{whalley+labbe1969} to extrapolate linearly to $213\K$. He also used the extrapolation to fill the gap at $\lambda \approx 170\ldots235\mum$ where no data were available.

Except for the treatment of the temperature dependence in \citet{warren1984}, only \citet{irvine+pollack1968} discussed the change in the absorption. However, their study only covers the $\lambda \leq 100\mum$ range.

A revised version of the data from \citet{warren1984} was published by \citet{warren+brandt2008}, who included more recent data, for example from \citet{curtis2005}. They used three measurements of \citet{curtis2005} at the highest temperatures ($156\K$, $166\K$ and $176\K$) and extrapolated their temperature dependence linearly to $266\K$. Unfortunately, the new data set offers no improvement at wavelengths $\lambda > 100\mum$, because of the lack of data in this region (see Fig. 6 in their publication).

\citet{li+greenberg1998} used a simpler way to create a new data set for amorphous water ice. They took the $10\K$ measurement from \citet{hudgins1993} for amorphous water ice up to $\lambda = 100\mum$ and extrapolated them with the aid of a simple power law ($\kappa \propto \lambda^{-0.5}$) up to $\lambda = 10\mm$. Although their model disagrees with some other measurements \citep[e.g.][]{whalley+labbe1969,mishima1983}, it remains the most commonly used data set for models of diverse astronomical objects, especially debris disks \citep[e.g.][]{su2005,loehne2012}.

\begin{figure}[t!]
  \centering
  \includegraphics[width=0.55\linewidth,angle=270]{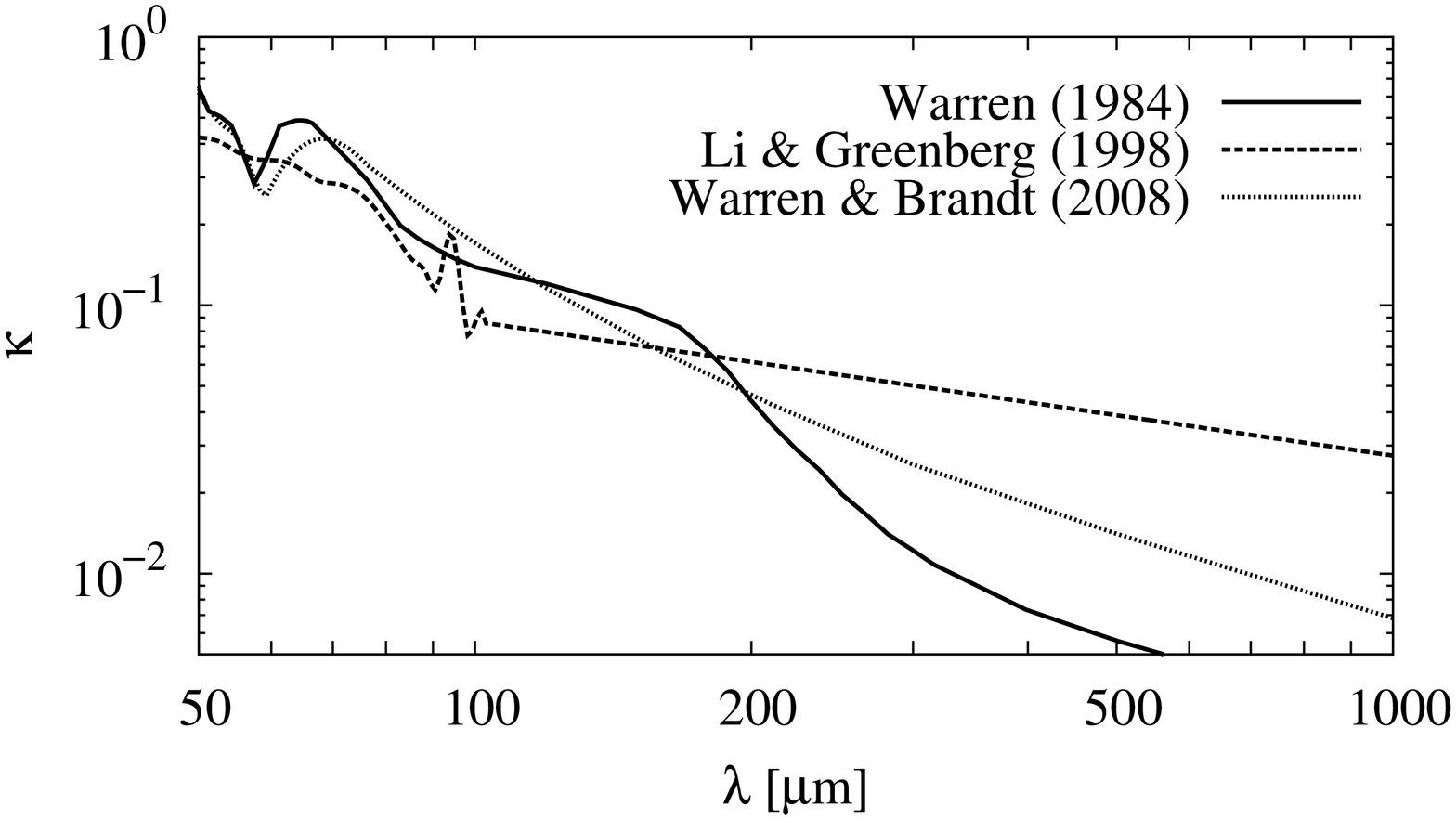}\\%
  \caption{Data for the imaginary part $\kappa$ of the refractive index $N$. Shown are the models from \citet{warren1984}, \citet{li+greenberg1998}, and \citet{warren+brandt2008}.
  \label{fig:literature_models}}
\end{figure}

All three models for ice in a range of $\lambda=50\ldots1000\mum$ are shown in Fig.~\ref{fig:literature_models}.
%
\section{Methods} \label{sec:methods}
\subsection{Experimental methods}
\subsubsection{Sample production}
The best controlable way to produce a layer of water ice 
is to deposit water vapour on a cooled substrate \citep[e.g.][]{smith1994,moore+hudson1992,hudgins1993,curtis2005}. In the far infrared, where $\kappa$ drops to $\sim 10^{-3}$ at $\sim 500\mum$ (see Fig.~\ref{fig:data_literature}), the absorption coefficient $\delta = 4\pi \kappa \lambda^{-1}$ is as small as $0.025\mm^{-1}$. Therefore, we need relatively thick ice layers to measure significant absorption. In this case, the depositing method is hard to use because it requires a very long deposition time.

Thicker ice layers can be obtained by freezing a certain amount of distilled water \citep[][]{whalley+labbe1969} or by producing a big single crystal of water ice \citep[][]{mishima1983}. Either way, this always results in ice of crystalline type.

\begin{figure}[t!]
  \centering
  \includegraphics[width=0.27\textwidth]{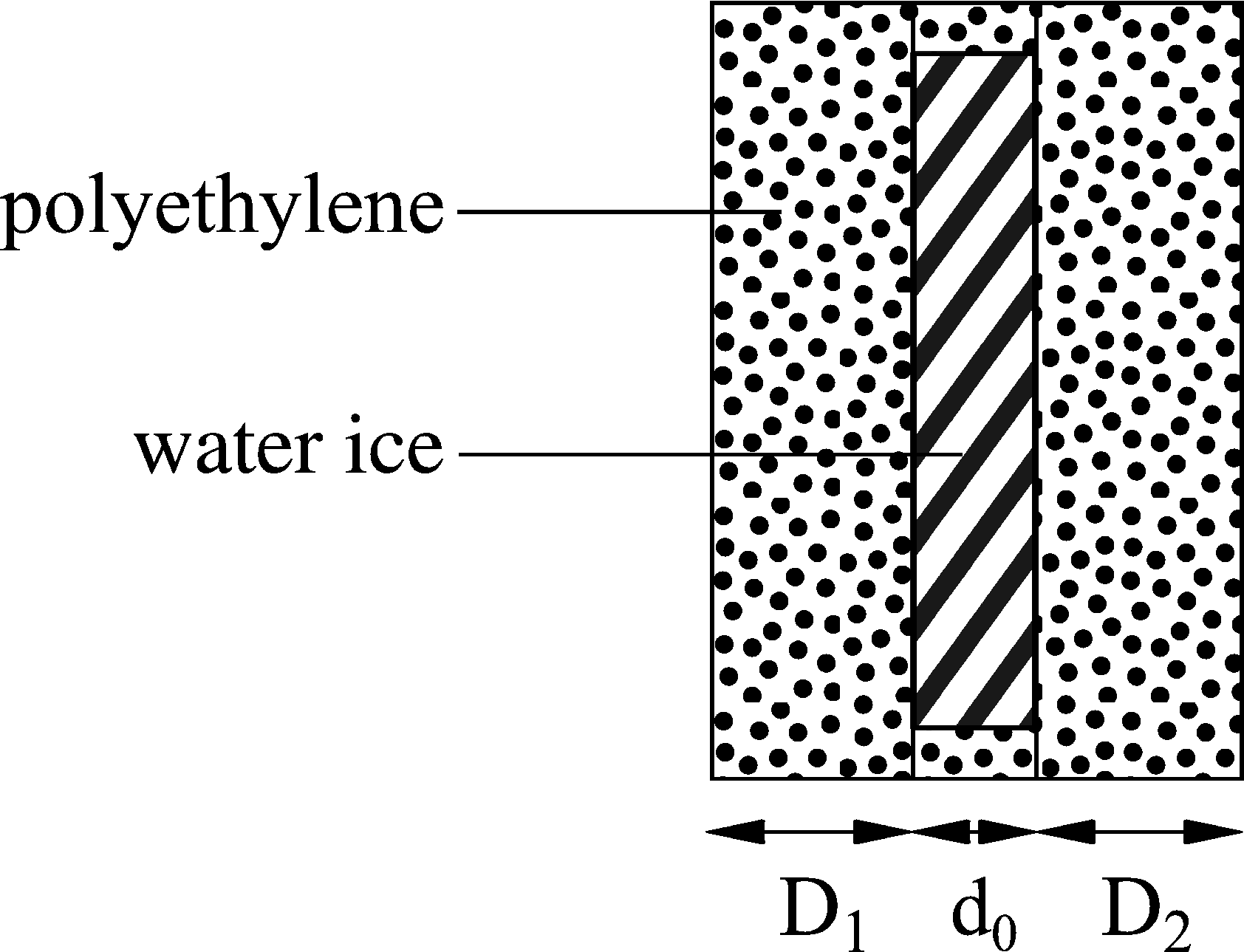}\\%
  \caption{Construction of the sample (details in the text).
  \label{fig:sample}}
\end{figure}

Our experiments are similar to those of \citet{whalley+labbe1969}. The samples were produced by putting double distilled water between two windows of polyethylene (see Fig.~\ref{fig:sample}) in a continuous flow helium cryostat (CryoVac KONTI Spektro B) and freezing it by cooling the cryostat down to $250\K$. One of the windows had an extra rim of height $d_0$. In this way, we obtained an ice layer of hexagonal ($I_h$) polycrystalline water ice with an approximate, nominal thickness $d_0$ for our measurements.

The windows were made from polyethylene powder that was pressed with a specially manufactured die at a pressure of $10\,\text{t} \cm^{-2}$. Afterwards the powder was fused at a temperature of $140^\circ C$. The resulting windows had a thickness of $D_{1,2}=0.91 \ldots 1.78\mm$ at room temperature. 
\subsubsection{Measurements}
All the measurements were taken in a Fourier transformation infrared spectrometer (BRUKER 113v) with the cryostat inserted into the sample chamber. We measured directly in the cryostat through four polyethylene windows in the walls of the cryostat. The inner board of the cryostat was cooled by liquid helium. With a contact gas (also helium) the sample was then cooled to 250\,K, 200\,K, 150\,K, 100\,K, 50\,K, and 10\,K. At each temperature we investigated the transmission $T$ by first taking a reference spectrum with an empty sample holder in the cryostat and then measuring the sample itself. By doing this we eliminated the influence of the optical devices on the spectrum of the sample.

We measured in three different wavelength regions and used two different detectors (one of deuterated triglycine sulfate and a LHe-cooled bolometer). The wavelength regions are mostly determined by the thickness of the mylar foil in the beam splitter. We used three different beam splitters with foils of $12\mum$, $23\mum$, or $125\mum$ where the thickest was used for the longest wavelengths.

Depending on the wavelength region, we varied $d_0$ from $100\mum$ to $3\mm$. To probe the shorter wavelengths with $\lambda = 45 \ldots 200\mum$, a thin layer of $d_0 \approx 100\mum$ was used (measurement 1, M1). For $\lambda = 100 \ldots 500\mum$, we used a layer of about $500\mum$ thickness (measurement 2, M2). For the longest wavelengths, $\lambda = 200 \ldots 1000\mum$, we chose $d_0 \approx 3000\mum$ (measurement 3, M3). The reason for this approach is the steepness of the absorption spectrum. Towards shorter wavelengths, the ice gets opaque, for longer wavelengths, the absorption drops below detectable values. How rapidly the drop turns out depends on the temperature $\vartheta$. Therefore, we chose the layer thickness in such a way that for representative wavelengths ($100\mum$, $200\mum$ and $350\mum$) of the corresponding spectral ranges the absorption does not drop below 20\% for all temperatures, but is still thin enough to be not opaque at these wavelengths. The range where reliable measurements 
are possible shifts to shorter wavelengths with decreasing thickness of the ice layer.
\subsection{The background spectrum}
As mentioned above, the sample consists not only of an ice layer, but also of two additional polyethylene windows. To isolate data describing only the ice, we needed to divide the measured transmission $T$ by a background spectrum $T_0$ that only derives from the absorption in the polyethylene windows.

When measuring such a background spectrum of the two polyethylene windows separated by a vacuum layer, the result is influenced by interference effects due to multiple reflection in the interlayer. These interferences almost disappear when the vacuum is replaced by ice later on, because the amplitude of the intereferences $A$ is proportional to $\Delta n^2$, where $\Delta n$ is the difference between the real parts of the reflective indices of polyethylene, $n_\text{PE}$, and the interlayer, $n_\text{vac/ice}$. The difference $\Delta n$ shrinks from $\Delta n \approx 1.5$ with vacuum to $\Delta n \approx 0.3$ with ice in between.
Because of these effects we had to remove the interferences in the background itself by taking only the polyethylene windows into account.

To do this, we used the commercial thin-layer optics program SCOUT 2\footnote{M.\,Theiss Hard- and Software, Aachen, Germany} to fit an oscillator model to the measured background spectrum. The model describes a three-layer system that consists of a polyethylene layer $D_1$, a vacuum layer $d_\text{vac}$, and another polyethylene layer $D_2$. The thickness of the vacuum interlayer $d_\text{vac}$ was considered unknown.

To describe the wavelength-dependent complex refractive index $N_\text{PE}(\tilde{\nu})=\sqrt{\varepsilon_\text{PE}(\tilde{\nu})}$ (with $\tilde{\nu}$ being the wavenumber and $\varepsilon$ the dielectric function) of polyethylene in this model we used two Lorentz oscillators, a and b. Oscillator b describes the absorption band of polyethylene near $130\mum$. The refractive index is then given by
\begin{equation}
N_\text{PE}(\tilde{\nu})^2 = \varepsilon_0 + \frac{\Gamma_a^2}{(\tilde{\nu}_{0,a}^2 - \tilde{\nu}^2)-i \tilde{\nu} \gamma_a} + \frac{\Gamma_b^2}{(\tilde{\nu}_{0,b}^2 - \tilde{\nu}^2)-i \tilde{\nu} \gamma_b}
\end{equation}
where $\tilde{\nu}_0$ is the resonance position, $\Gamma$ is the strength, $\gamma$ is the damping of an oscillator, and $\varepsilon_0$ was set to $2.25$ ($\varepsilon_0 \lesssim n_\text{PE}(\tilde{\nu})^2$).
For the vacuum layer, $N_\text{vac}(\tilde{\nu})^2 = \varepsilon_\text{vac}(\tilde{\nu}) = 1$ applies.

\begin{figure}[t!]
  \centering
  \includegraphics[width=0.5\linewidth,angle=270]{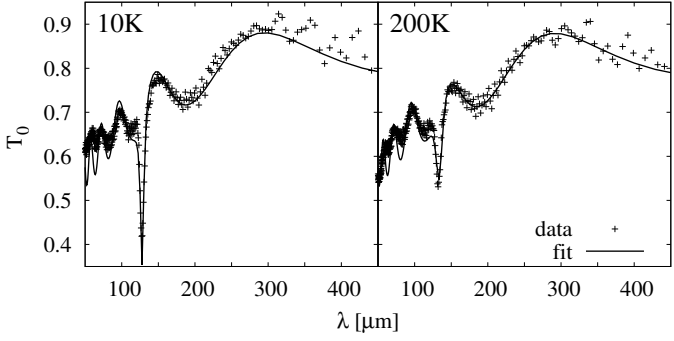}\\%
  \caption{Examples for the transmission $T_0$ of the background measurement and the corresponding oscillator fit. For the background measurement the sample only consisted of the two polyethylene windows with $D_1$ and $D_2$ (here: $D_1 = 1.68\mm$ and $D_2 = 1.65\mm$) and a vacuum interlayer of thickness $d_\text{vac}$ (best-fit here: $d_\text{vac}=143\mum$).
  \label{fig:background-fit}}
\end{figure}

Varying the parameters $\tilde{\nu}_0$, $\Gamma$, and $\gamma$ for each of the two oscillators as well as the thickness of the vacuum layer $d_\text{vac}$, the transmission calculated from the oscillator model was fitted to the measured background spectrum. Figure~\ref{fig:background-fit} demonstrates that the model reproduces the data quite well. There is only a slight discrepancy at the shorter wavelengths. It traces back to the divergence of the beam of the spectrometer that limits its coherence at short wavelengths. This effect was not considered in the model.

\begin{figure}[t!]
  \centering
  \includegraphics[width=0.5\linewidth,angle=270]{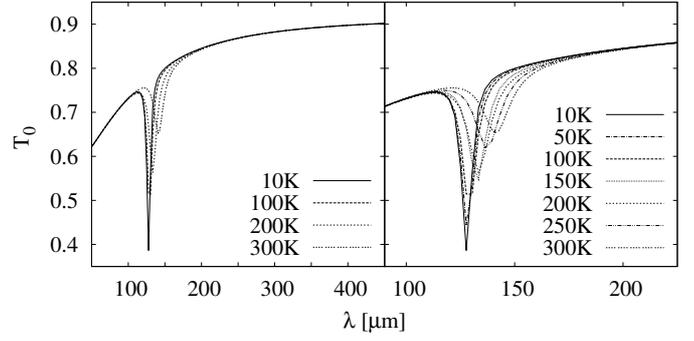}\\%
  \caption{Calculated background spectra $T_0$ for a system of two polyethylene windows with $D_1 = 1.68\mm$ and $D_2 = 1.65\mm$ at temperatures between 10\,K and 250\,K. On the right is a close up on the region of the feature and the temperature dependence therein.
  \label{fig:background-result}}
\end{figure}

Once all oscillator parameters for polyethylene and the (best fitting) thickness of the vacuum interlayer are known, the vacuum layer can be removed from of the model. Then, with the found $N_\text{PE}(\tilde{\nu})$, we were able to calculate a transmission spectrum for two polyethylene windows, even with variable thicknesses $D_1$ and $D_2$. This way we got a modelled background spectrum $T_0$ for each sample configuration and temperature. For exemplary thicknesses $D_1$ and $D_2$, the resulting background spectra $T_0$ for different temperatures $\vartheta$ between 10\,K and 250\,K are shown in Fig.~\ref{fig:background-result}.

Clearly, the background spectrum is temperature dependent. There is a shift to longer wavelengths and a broadening of the $130\mum$ feature of polyethylene. Accordingly, the parameters for oscillator b had to be calculated for all temperatures used in the measurements. In contrast, oscillator a, which causes the continuum absorption, showed no significant temperature dependence after all, having a constant $\gamma$ over the entire temperature range. All calculated parameters for a and b are listed in Table~\ref{tab:oscillator-parameters}. Note that the resonance position $\tilde{\nu}_0$ is almost constant, while the actual feature position in the spectra $\tilde{\nu}'_0 = \left( \tilde{\nu}_0^2-\frac{1}{2}\gamma^2 \right)^\frac{1}{2}$ is temperature dependent, since $\gamma = \gamma(\vartheta)$.

\begin{table}[t!]
  \centering
  \caption{Best-fit oscillator parameters $\Gamma$, $\tilde{\nu}_0$ (or $\tilde{\nu}_0'$) and $\gamma$ for the modelled background spectra at all measuring temperatures.}
  \label{tab:oscillator-parameters}
  
  \begin{tabular}[h!]{cc|cccc}
  \hline
oscil.    & $\vartheta$   & $\Gamma$ [$\cm^{-1}$] & $\tilde{\nu}_0$ [$\cm^{-1}$] & $\gamma$ [$\cm^{-1}$] & $\tilde{\nu}_0'$ [$\cm^{-1}$] \\
		\hline
		a &         & $28.7$  & $4.0 \times 10^6$ & $5.7 \times 10^6$ & $514.4$ \\
		\hline
		b & $10\K$  & $1.19$  & 78.4 & ~$6.8$  & $78.3$\\
		  & $50\K$  & $1.21$  & 78.4 & $10.9$  & $78.0$\\
		  & $100\K$ & $1.19$  & 78.2 & $15.2$  & $77.5$\\
		  & $150\K$ & $1.15$  & 77.3 & $24.1$  & $75.4$\\
		  & $200\K$ & $1.18$  & 80.2 & $41.0$  & $74.8$\\
		  & $250\K$ & $1.10$  & 79.3 & $44.9$  & $72.8$\\
		  & $300\K$ & $1.02$  & 80.1 & $53.3$  & $70.7$\\
		\hline
  		\end{tabular}
\end{table}
\subsection{Correction of Data}
In each of the three wavelength ranges measured, the transmission of the respective ice layer ($100\mum$, $500\mum$, or $3\mm$) varied between very low values and almost complete transparency. The strong overlap of the wavelength ranges allowed us to use only those measured values where the absorption was detected with the highest sensitivity and accuracy, i.\,e. in the range of transmission values between 10\% and 90\% of the maximum value. The wavelength ranges resulting from this constraint are listed in Table~\ref{tab:wavelength_ranges}.

\begin{table}[t!]
  \centering
  \caption{Used wavelength ranges for the measurements M1 to M3.}
  \label{tab:wavelength_ranges}
  
  \begin{tabular}[h!]{c|ccc}
\hline
$\vartheta$ & M1 & M2 & M3 \\
\hline
~$10\K$ & $(80-192)\mum$ & $(149-313)\mum$ & $(225-370)\mum$ \\
~$50\K$ & $(82-192)\mum$ & $(153-323)\mum$ & $(225-400)\mum$ \\
$100\K$ & $(82-185)\mum$ & $(154-345)\mum$ & $(229-417)\mum$ \\
$150\K$ & $(85-182)\mum$ & $(162-345)\mum$ & $(240-500)\mum$ \\
$200\K$ & $(87-182)\mum$ & $(171-370)\mum$ & $(259-556)\mum$ \\
$250\K$	& $(93-192)\mum$ & $(182-385)\mum$ & $(296-625)\mum$ \\
\hline
  \end{tabular}
\end{table}
The comparison with M1 and M3 showed that M2 suffered from a relatively strong additional intensity loss caused by scattering effects and deflection of the beam. This can be caused by inhomogeneities of the ice layer or impurities in the ice itself. Since for the thinner ice layers the layer thickness is also more uncertain, we decided to normalize M1 and M2 to the values of M3. The latter showed almost no additional losses and the layer thickness should be accurate to about $10\%$. The only correction done for M3 was to include the reflection losses at the two polyethylene-ice boundaries of the sample, which were not considered in the model of the background spectra by raising the transmission by about $1.8\%$.

Intensity loss, on the one hand, and the uncertainty in the thickness, on the other hand, suggest using two normalization factors. First, the transmission is corrected by a factor $f_\text{corr}$ giving a higher transmission $T_{\text{corr},j} = (T_j/T_{0,j}) \times f_\text{corr}$, with $T_j=T(\lambda_j)$ being the original measured data points and $T_{0,j}=T_0(\lambda_j)$ the background transmission at the same wavelengths $\lambda_j$. Second, we varied the nominal thickness $d_0$ to a corrected value of $d_\text{corr}$. We determined the values for the correction by minimizing the difference of the measured transmissions in the overlapping wavelength regions first of M3 and M2, and second of the corrected version of M2 and the original M1. The correction factors were determined for each temperature and then averaged to get a single set for all temperatures.

As a result, the nominal thickness $d_0=100\mum$ for M1 turned out to require a stronger correction to $d_\text{corr}=202 \pm 33 \mum$, whereas the measured transmission of M1, in principle, needed no correction. For M2 the opposite was found: the nominal thickness $d_0 = 500\mum$ was accurate to 10\%, whereas the transmission had to be corrected by a factor of $f_\text{corr}=1.6 \pm 0.1$.
\subsection{Describing the absorption}
To describe the absorption, we chose the imaginary part $\kappa$ (many other terms are used in the literature) of the complex refractive index $N$. As explained before, this is most appropriate, because $\kappa$ is a constant of the material itself and therefore independent of the geometry of the sample.

To gain $\kappa(\lambda_j)$ from the transmission measurements at each wavelength $\lambda_j$, we applied the equation
\begin{equation}
\kappa(\lambda_j) = -\text{ln}(T_{\text{corr},j}) \times \left( \frac{\lambda_j}{4\pi} \right) \times \left(\frac{1}{d_\text{corr}} \right)
\end{equation}
to all points of the transmission measurement $T_j$.
%
\section{Results} \label{sec:results}
%
\begin{figure}[t!]
  \centering
  \includegraphics[width=0.96\linewidth,angle=270]{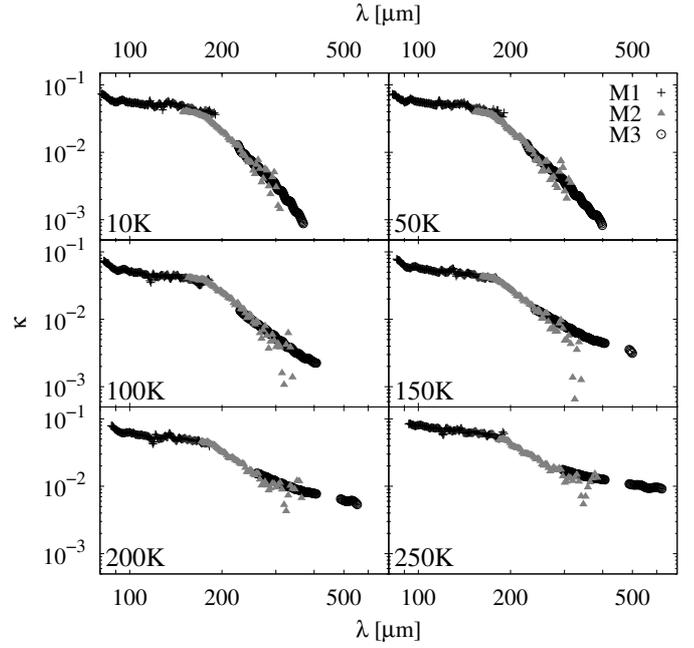}\\
  \caption{Coefficient $\kappa$ for the measurements M1 to M3 at the temperatures $10\K$, $50\K$, $100\K$, $150\K$, $200\K$ and $250\K$ for $\lambda \gtrsim 100\mum$.
  \label{fig:measured_data}}
\end{figure}
In Fig.~\ref{fig:measured_data} the resulting data are displayed for the three measurements M1 to M3 at all six temperatures from $10\K$ to $250\K$. The measurement M1 yields a smooth curve with only a bit of noise at the longer wavelength end of its domain. In some cases some artefacts appear that are caused by incomplete reduction of the $130\mum$ feature of polyethylene in the background spectrum. The data from M2 are much noisier, especially at higher temperatures and at longer wavelengths, whereas those from M3 are very smooth again, with only slight interference effects (fringe pattern) at wavelengths beyond $500\mum$. Besides, a low beam intensity around $447\mum$ caused by the optical properties of the beam splitter ($125\mum$ polyethylene foil) forced us to leave out the wavelengths from $416.7\mum$ to $476.2\mum$ in M3.
\subsection{Temperature dependence}
The behaviour of $\kappa$ obviously changes at around $\lambda=200\mum$. Short-ward of that wavelength, the slope stays nearly constant. At longer wavelengths, $\kappa$ decreases with increasing $\lambda$, and there is a clear temperature dependence of the slope.
To quantify these visual impressions, we fitted the data with power laws of the form
\begin{equation}
\kappa = \varphi \times \left( \frac{\lambda}{\lambda_0} \right) ^{-\alpha},
\end{equation}
where $\varphi$ and $\alpha$ are constant for all temperatures in each of the three measurements M1 to M3. The parameter of interest is $\alpha$ because it basically describes the slope of $\kappa$. The constant $\lambda_0$ is used for normalization and was set to $200\mum$ as a typical wavelength.

\begin{figure}[t!]
  \centering
  \includegraphics[width=0.7\linewidth,angle=270]{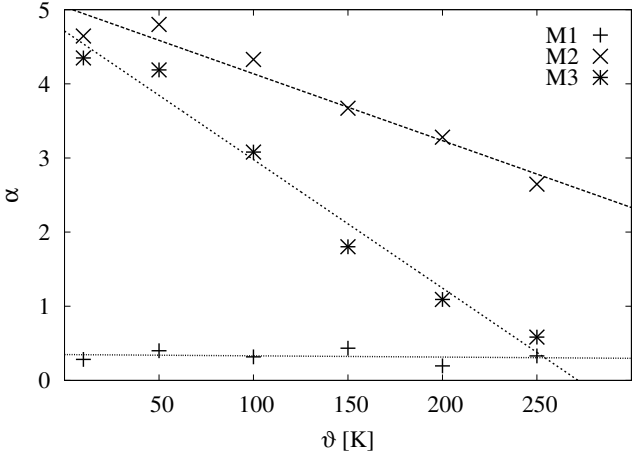}\\%
  \caption{Slopes $\alpha$ of $\kappa(M1)$ to $\kappa(M3)$ depending on the temperature $\vartheta$. The lines indicate the linear interpolation.
  \label{fig:slope}}
\end{figure}

We fitted the power law at the three measurements 1 to 3, which lead us to the fits $\kappa(M1)$ to $\kappa(M3)$. The fitting was done in the central 50\% of the ranges listed in Table~\ref{tab:wavelength_ranges}. The resulting $\alpha$ depending on the temperature is plotted in Fig.~\ref{fig:slope}.

\begin{figure}[t!]
  \centering
  \includegraphics[width=0.7\linewidth,angle=270]{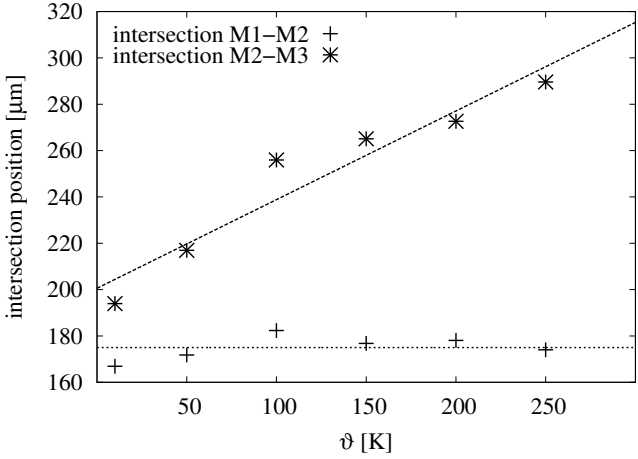}\\%
  \caption{Wavelength of intersection between $\kappa(M1)$ and $\kappa(M2)$, and $\kappa(M2)$ and $\kappa(M3)$ at different temperatures $\vartheta$. The lines indicate the linear interpolation.
  \label{fig:intersection}}
\end{figure}

The slope of $\kappa(M1)$ is flat, $\alpha_1 = 0.1 \ldots 0.5$, over the entire temperature range, whereas that of $\kappa(M2)$ and $\kappa(M3)$ is much steeper. For $\kappa(M2)$, $\alpha$ is decreasing with temperature from $\alpha_2 \simeq 5$ ($\vartheta=10\K$) to $\alpha_2 \simeq 2.5$ ($\vartheta=250\K$). An even stronger temperature dependence occurs for $\kappa(M3)$, with $\alpha$ ranging from more than 4 at $10\K$ to about only $0.5$ at $250\K$. A linear fitting with $\alpha_i (\vartheta)= m_i \times \vartheta + n_i$ ($i=1,2,3$) results in
\begin{eqnarray}
m_1 &\approx & -0.0004 \K^{-1},
\label{equ:linear_slopes1}\\
m_2 &=& (-0.009 \pm 0.001) \K^{-1} \text{, and}
\label{equ:linear_slopes2}\\
m_3 &=& (-0.017 \pm 0.001) \K^{-1}.
\label{equ:linear_slopes3}
\end{eqnarray}
To determine more accurately the wavelength where the behaviour of the slope is changing, we searched for the intersection of the three power law fits at each temperature $\vartheta$ (see Fig.~\ref{fig:intersection}).

The intersection point between $\kappa(M1)$ and $\kappa(M2)$ is nearly independent of temperature, $\lambda_\text{intersec. 1-2} = (175 \pm 2)\mum$. In contrast, the intersection between $\kappa(M2)$ and $\kappa(M3)$ depends on temperature linearly:
\begin{equation}
\lambda_\text{intersec. 2-3} = (0.38 \pm 0.06)\mum \K^{-1} \times \vartheta + (201 \pm 9)\mum.
\label{equ:temp_dependecy_position}
\end{equation}
Such a behaviour can be explained by a spectral feature that is broadening with temperature, and therefore the slope $\alpha$ is decreasing with temperature in the region of the long wavelength wing of the feature. In fact, \citet{wozniak+dera2007} reported an absorption band at around $166\mum$ caused by an intermolecular bending mode that can be associated with the intersection value of $\kappa(M1)$ and $\kappa(M2)$.
\subsection{Analytic model}
\begin{figure}[t!]
  \centering
  \includegraphics[width=0.55\linewidth,angle=270]{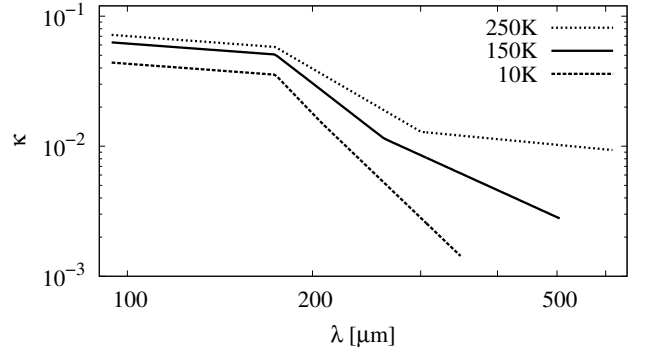}\\%
  \caption{The analytic model for $\kappa$ calculated for the temperatures 10\,K, 150\,K, and 250\,K (formulas are given in the text).
  \label{fig:data_set_examples}}
\end{figure}

Having determined the temperature dependence, we now create a simple analytic model that can be applied to the actual temperature of any object of interest. For this purpose, we assume three power laws
\begin{equation}
\kappa_i(\varphi_i,\alpha_i) = \varphi_i \times \left( \frac{\lambda}{\lambda_0} \right) ^{-\alpha_i},
\qquad (i=1,2,3) 
\label{equ:model}
\end{equation}
to account for the three distinct regions with different behaviour of the absorption. Region 1 starts at $\lambda = 94\mum$ right after the strong absorption features of water ice at around $44\mum$ and $63\mum$. The other two regions are separated by the intersection wavelengths $\lambda_\text{intersec. 1-2}$ and $\lambda_\text{intersec. 2-3}$ found above. The longest wavelength $\lambda_\text{max}$ covered by the model is the maximum wavelength probed in measurement 3 (see Table~\ref{tab:wavelength_ranges}). It increases linearly with temperature:
\begin{equation}
\lambda_\text{max} = 1.1 \times \vartheta + 341\mum.
\end{equation}
The slopes $\alpha_i$ are those from the linear fits $\alpha_i(\vartheta)$ described above (equations~\ref{equ:linear_slopes1} to~\ref{equ:linear_slopes3}):
\begin{eqnarray}
\alpha_1 &=& const. = 0.35, \\
\alpha_2 &=& - 0.009 \times \vartheta + 5.0 \text{, and}\\
\alpha_3 &=& - 0.017 \times \vartheta + 4.7.
\end{eqnarray}
The factors  $\varphi_i$ were determined by finding the best possible fitting at the region boundaries. The result was a nearly constant value for $\varphi_1$ and a linear dependence for $\varphi_2$ and $\varphi_3$:
\begin{eqnarray}
\varphi_1 &=& 0.0188,\\
\varphi_2 &=& 0.017 + (9 \times 10^{-5}) \times \vartheta \text{, and}\\
\varphi_3 &=& 0.034 + (8.9 \times 10^{-5}) \times \vartheta.
\label{equ:phi_3}
\end{eqnarray}
Note that the value for $\varphi_1$ is somehow artificial and due to the fitting at the region boundaries. There is no clear temperature dependence like this seen in the measurements. However, the resulting model from equations \ref{equ:model}--\ref{equ:phi_3} is visualised in Fig.~\ref{fig:data_set_examples} for three exemplary temperatures.

Further, to make the model usable, we had to extend the data set to $\lambda < 94\mum$ and $\lambda > \lambda_\text{max}$. We did this by applying the $\lambda=15 \ldots 94\mum$ measurements from \citet{curtis2005} for crystalline ice at a temperature of $136\K$. At $\lambda < 15\mum$ we used data from \citet{li+greenberg1998}. Finally, at $\lambda_\text{max}< \lambda < 10\mm$ the absorption was computed by an extrapolation of $\kappa_3(\varphi_3,\alpha_3)$.

\begin{figure}[t!]
  \centering
  \includegraphics[width=0.49\linewidth,angle=270]{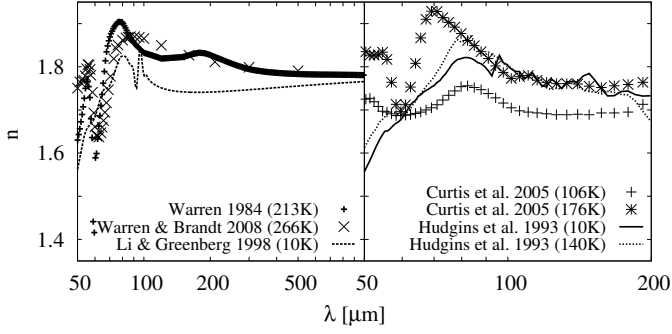}\\%
  \caption{Real part $n$ of the complex refractive index $N$ for different data sets.
  \label{fig:real_index}}
\end{figure}

For most applications, the real part $n$ of the complex refractive index $N$ is also needed. The Kramers-Kronig relation links $n$ and $\kappa$, which means that a similarity in $\kappa$ of two data sets implies a similarity in $n$ of the two. But contrary to $\kappa$, $n$ does not spread over a wide range in the far infrared in the data sets discussed above. In case of crystalline water ice $n$ always stays in a narrow range of $n=1.75\ldots1.85$ with no strong temperature dependence, for $\lambda \geq 100\mum$ (see Fig.~\ref{fig:real_index}). Amorphous ice has a slightly reduced refractive index (see data of Li \& Greenberg or Curtis at $106\K$). Minor temperature related variations seem to be linked to the change of the $44$ and $63\mum$ features in the absorption. Therefore, we just used temperature independent data of \citet{warren1984} for $n$ in our model in the wavelength range $\lambda>94\mum$. This compilation used, amongst others, the data of \citet{mishima1983} and \citet{whalley+labbe1969}, which have good agreement with our data in terms of $\kappa$ (see section~\ref{sec:comparison_literature}).
At shorter wavelengths ($\lambda<94\mum$) we used data from \citet{curtis2005} and \citet{li+greenberg1998} for $n$ in the same wavelength ranges as for $\kappa$.
%
\section{Discussion} \label{sec:discussion}
\subsection{Comparison with literature}
\label{sec:comparison_literature}
\begin{figure}[t!]
  \centering
  \includegraphics[width=0.88\linewidth,angle=270]{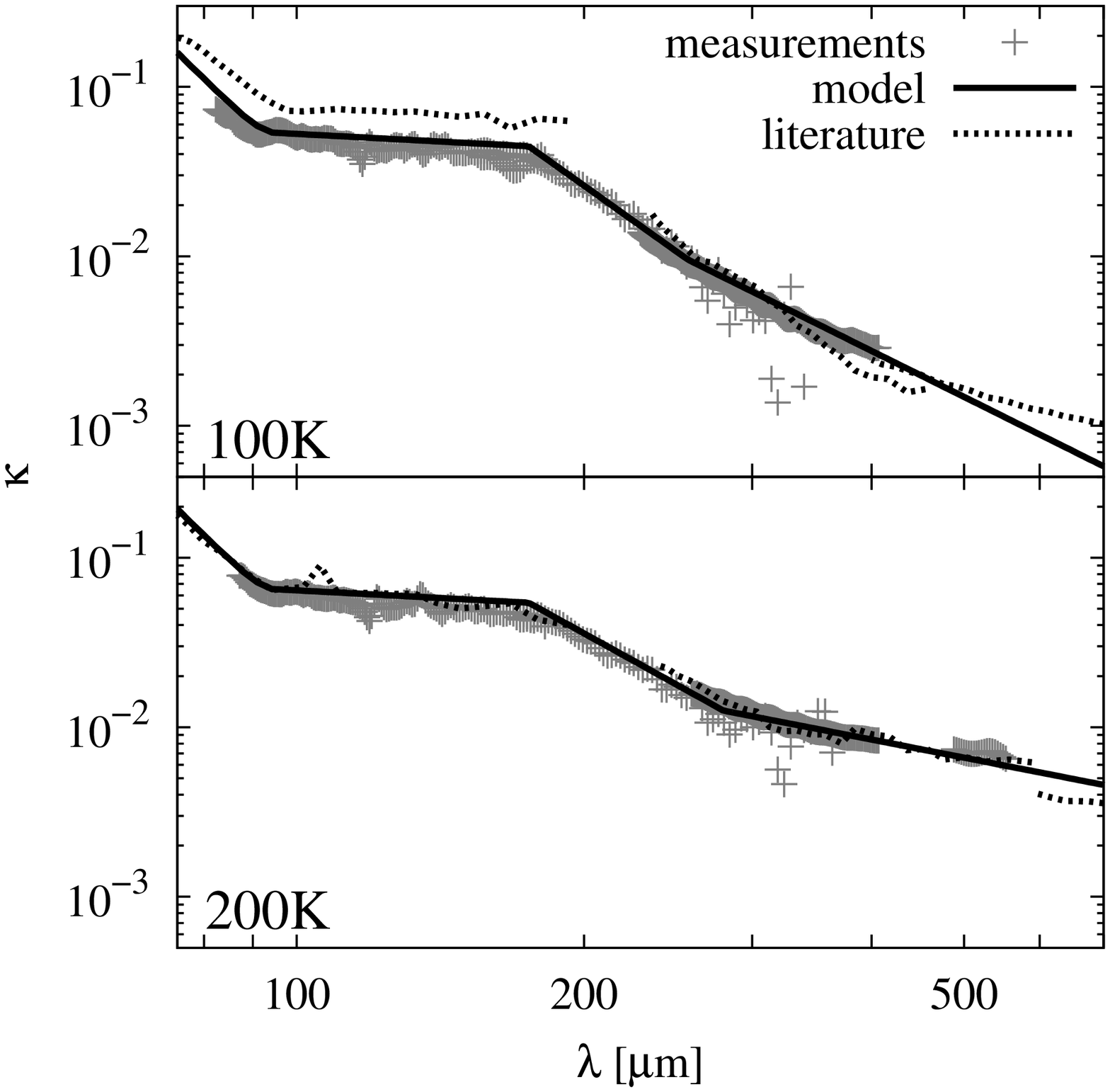}\\%
  \caption{Comparison of literature data for $\kappa$ and our measurements at temperatures of $100\K$ (top) and $200\K$ (bottom). Literature for $100\K$: \citet{curtis2005} at $106\K$, \citet{whalley+labbe1969} at $100\K$, \citet{mishima1983} at $100\K$. For $200\K$: \citet{curtis2005} at $176\K$, \citet{whalley+labbe1969} at $200\K$, \citet{mishima1983} at $200\K$. All literature data are for crystalline ice, except for the $106\K$-measurement of \citet{curtis2005}, which is for amorphous ice.
  \label{fig:data+literature}}
\end{figure}

We compared both our measured data and our model with literature data for the two temperatures $100\K$ and $200\K$. We chose the data  sets of \citet{curtis2005}, \citet{mishima1983}, and \citet{whalley+labbe1969} for this comparison and generally found good agreement with our data (see Fig.~\ref{fig:data+literature}), with a few exceptions reviewed below. Also, our model reproduces our measured data with good accuracy.

There is a slight disagreement between our measurements and those by \citet{curtis2005} at $106\K$, which could be caused by the fact that their data at this temperature are for amorphous ice, since amorphous ice seems to have a higher absorption in the region right after the strong feature (see Fig.~\ref{fig:vgl_curtis}). In addition, a small peak in their $176\K$ measurement at about $100\mum$ is not seen in our data. Note, however, that this peak was not found in any other measurement from the literature either. At $100\K$, the data of \citet{mishima1983} suggest a slightly higher absorption compared to our results. This may possibly be caused by extrapolating $\kappa_3(\varphi_3,\alpha_3)$ to $\lambda > \lambda_\text{max}$, where it is not supported by measurements.

\begin{figure}[t!]
  \centering
  \includegraphics[width=0.7\linewidth,angle=270]{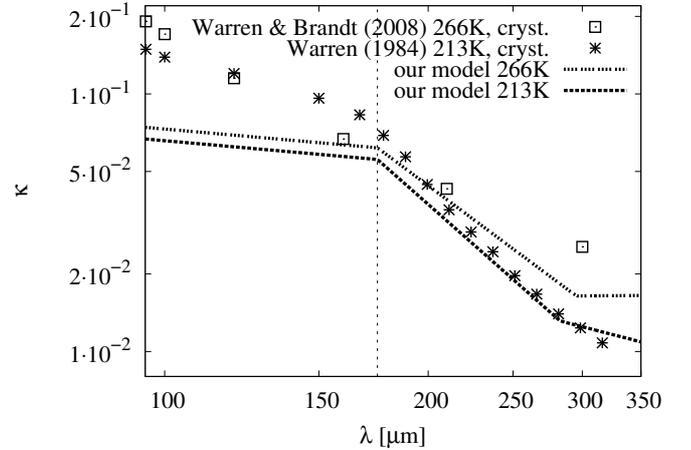}\\%
  \caption{Coefficient $\kappa$ in the region around the $175\mum$ feature in the data sets of \citet{warren1984} and \citet{warren+brandt2008} compared to our model at the same temperatures.
  \label{fig:166feature_literature}}
\end{figure}

The feature that we found at $175\pm6\mum$ can also be seen in the compilation data of \citet{warren1984} with the caveat that the underlying data have a gap surrounding this wavelength. Yet, \citet{warren1984} noticed that the data sets at different temperatures converge at a wavelength of around $166\mum$. Although he offered no explanation, why this is happening, he based his interpolation in this wavelength range on this fact. As a result, he obtained a slight bend in the absorption index $\kappa$ (see Fig.~\ref{fig:166feature_literature}). Later, \citet{wozniak+dera2007} associated this bend with an absorption band of crystalline water ice. They argued that the feature is caused by intermolecular bending modes. In the revised publication of Warren's data \citep{warren+brandt2008}, the feature is not seen at all because of the scarcity of data points in this area (see Fig.~\ref{fig:166feature_literature}).
\subsection{Outlook}
What still remains largely unexplored are the possible fine differences in optical properties of different types of ice. Although Fig.~\ref{fig:vgl_curtis} suggests that most of the differences between amorphous and crystalline ice are in the regions of spectral features, this is only valid at $\lambda \leq 200\mum$. We do not know the behaviour of amorphous ice at longer wavelengths. For silicates, there is an immense rise in the absorption for the amorphous structure at long wavelengths \citep{fabian2000}. This is because of the disorder of the solid that allows for absorption by vibration modes that cannot be excited in an ordered crystal. The same effect could possibly occur for water ice. To verify this, it would be straightforward to carry out measurements similar to those described here, but with amorphous ice. However, we cannot produce amorphous ice by freezing distilled water. We would have to use the depositing method, which is not easy, as sufficiently thick layers are required for reliable measurements.

Besides possible differences between amorphous and crystalline ice there might be differences in the crystalline ice itself, since two possible types (hexagonal ice $I_h$ and cubic ice $I_c$) exist in astronomical environments. Also, the knowledge about differences in polycrystalline and monocrystalline ice needs significant improvement. Future measurements should address these uncertainties as well.
%
%
\section{Application on the debris disk HD\,207129} \label{sec:application}
Of many possible applications of our model, we have selected here one specific object to illustrate how our model works and which kind of implications replacing one model of ice with another might have.

The object in question is the debris disk around the G2V star HD\,207129 at a distance of 15.6\,pc \citep{rhee2007}. The disk has been observed by many instruments at wavelengths from mid-infrared through submillimetre wavelengths, resulting in a densely sampled spectral energy distribution (SED) and resolved images at several wavelengths. Accordingly, detailed models of the disks have been developed. For more information about the object, the observational data, and models that describe the disk, see \citet{marshall2011,loehne2012} and references therein.

\begin{figure}[t!]
  \centering
  \includegraphics[width=1.0\linewidth,angle=270]{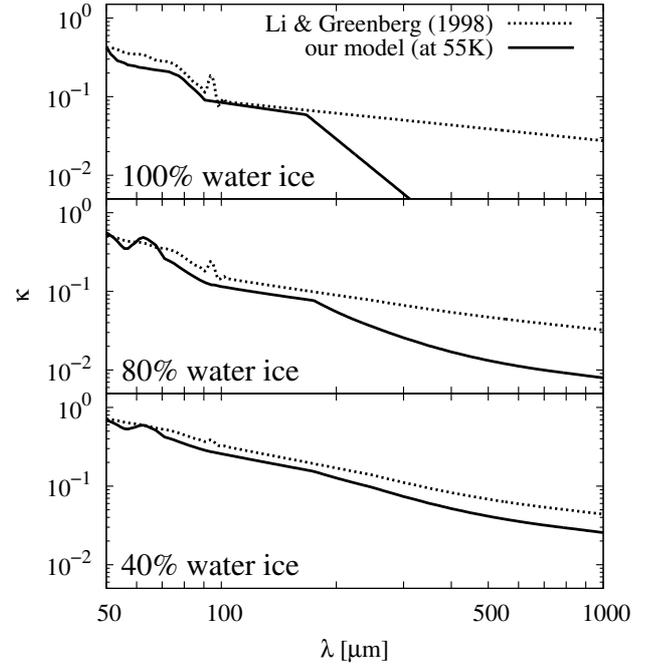}\\%
  \caption{Coefficient $\kappa$ in our model (at $\vartheta=55\K$) and in the data of \citet{li+greenberg1998} for pure ice (top), a mixture of 80\% ice and 20\% astronomical silicate (middle), and 60\% ice and 40\% astronomical silicate (bottom).
  \label{fig:comparison_ice-models}}
\end{figure}

It is not our intention to create a more accurate model of the debris disk than was done previously.  We merely aim to show the principal impact of the ice data on the disk's SED. Therefore, we chose a very simple model consisting of one dust ring and one material component. The disk was described by a power law model of the form
\begin{equation}
n(r,s) \propto r^{-\xi} \times s^{-\eta},
\end{equation}
which gives the number density $n(r,s)$ at each radius $r \in [r_\text{min},r_\text{max}]$ and particle size $s \in [s_\text{min},s_\text{max}]$. Both $\xi$ and $\eta$ were assumed to be constant over the entire disk. The values for all the parameters were taken from the modelling paper of \citet{loehne2012}:
\begin{itemize}
\item[] $r_\text{min} = 57\au$, $r_\text{max} = 194\au$, $\xi = -2.2$,
\item[] $s_\text{min} = 2.8\mum$, $s_\text{max} = 1\mm$, and $\eta = 3.8$.
\end{itemize}
These values were found for a dust composition with 50\% water ice (data from  \citet{li+greenberg1998}) and 50\% astro-silicate (data from \citet{draine2003}). In our simulations of the SED\footnote{For the simulation we used to the program "SEDUCE" (short for "SED Utility For Circumstellar Environments") from \citet{mueller2010}.}, we assumed a dust composition with different fractions of ice ranging from 40\% to 80\% with the optical data given by our own model (at a temperature of 55\,K). For comparison, we did the same with the data given by \citet{li+greenberg1998}. Similar to \citet{loehne2012}, we mixed the optical constants of ice and astronomical silicate \citep{draine2003} by means of the \citet{bruggeman1935} mixing rule. The absorption index $\kappa$ for the two (pure) ices and two exemplary mixtures is shown in Fig.~\ref{fig:comparison_ice-models}. One can see a clear dominance of the astronomical silicate on the optical properties of the ice-silicate mixture, because the difference of a mixture 
with 40\% and 80\% ice is relatively small compared with 80\% ice to 100\% ice. Note that the optical properties of the astronomical silicate are also uncertain because the material was first introduced to describe the observations in the interstellar medium (ISM) \citep{draine+lee1984}. Since the ISM is a completely different environment than the neighbourhood of a star, the optical properties of the silicate dust might be different here.

%
Strictly speaking, we are not allowed to use the parameters from \citet{loehne2012} for ice fractions other than 50\%. In principle, we would need to vary all of the parameters simultaneously, trying to find the best fit (with a least $\chi^2$-test) to the observational data (given in Table~\ref{tab:data_HD207129}) for each of the mixtures and each of the ice models. However, for illustrative purposes we decided to use a simpler approach and varied only one parameter, $\eta$, being aware that a change of $\eta$ implies a change in other disk parameters, too. 

\begin{table}[h!]
\centering
\caption{Photometry data of HD207129 for $\lambda > 20\mum$}
\label{tab:data_HD207129}
\begin{tabular}[h!]{lll}
\hline	
$\lambda$ [$\mum$] & $F_\lambda^\text{tot}$ [mJy] & Instrument (reference) \\
\hline
~24 & $155 \pm 5.3$ & MIPS \citep{trilling-et-al-2008}\\
~32 & $111 \pm 5.1$  & IRS \citep{krist2010}\\
~60 & $228 \pm 34$  & IRAS (Faint Source Catalogue)\\
~60 & $291 \pm 58$ & ISOPHOT\\
    &              & \citep{jourdain-de-muizon-et-al-1999}\\
~70 & $284 \pm 29$  & PACS \citep{marshall2011}\\
~70 & $278 \pm 11$ & MIPS \citep{trilling-et-al-2008}\\
~90 & $283 \pm 57$  & ISOPHOT\\
    &              & \citep{jourdain-de-muizon-et-al-1999}\\
100 & $311 \pm 36$ & PACS \citep{marshall2011}\\
160 & $211 \pm 42$ & PACS \citep{marshall2011}\\
160 & $250 \pm 40$ & MIPS \citep{krist2010}\\
160 & $150 \pm 20$ & MIPS \citep{tanner2009}\\
250 & $113 \pm 18$ & SPIRE \citep{marshall2011}\\
350 & $44.3 \pm 9$ & SPIRE \citep{marshall2011}\\
500 & $25.9 \pm 8$  & SPIRE \citep{marshall2011}\\
870 & ~~~~$5 \pm 3$ & APEX \citep{nilsson2010}\\
\hline
\end{tabular}
\end{table}
For the photosphere of the star we used a model from the Phoenix NextGen grid of \citet{hauschildt1999} that corresponds to the temperature, metallicity, and surface gravity of the star. According to \citet{rhee2007}, the disk has a black-body temperature of $55\K$, which we used as $\vartheta$ in our ice model.

\begin{figure*}[t!]
  \centering
  \includegraphics[width=0.3\linewidth,angle=270]{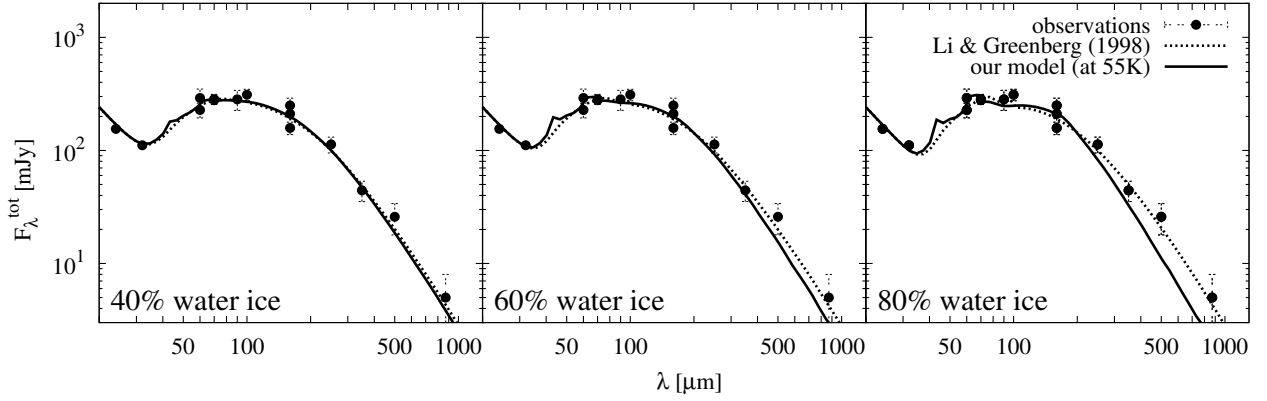}\\%
  \caption{Modelled SED of the debris disk around HD\,207129. From left to right: with 40\%, 60\%, and 80\% of water ice in the dust material (the rest is astronomical silicate from \citet{draine2003}). Compared are two models for the optical data of water ice: our model (at $\vartheta=55\K$) and the model of \citet{li+greenberg1998}.
  \label{fig:HD207129}}
\end{figure*}

The resulting SEDs of the simulations for three amounts of ice are shown in Fig.~\ref{fig:HD207129}. The best-fit $\eta$ values for five ice-silicate mixtures and each of the two ice data sets are listed in Table~\ref{tab:best-fit_eta}.

\setlength{\tabcolsep}{1.5mm}
\begin{table}[h!]
\centering
\caption{The best-fit $\eta$, the dust mass $m_\text{dust}$, and corresponding reduced $\chi^2$ for each fraction of water ice.}
\label{tab:best-fit_eta}
\begin{tabular}[h!]{l|lll|lll}
\hline
fraction & \multicolumn{3}{c|}{our model}                                & \multicolumn{3}{c}{model of Li \& } \\	
of water & \multicolumn{3}{c|}{($\vartheta=55\K$)}                       & \multicolumn{3}{c}{Greenberg (1998)} \\
ice      & $\eta$ &$m_\text{dust}$ [$M_{\oplus}$] & $\chi^2_\text{red.}$ & $\eta$ & $m_\text{dust}$ [$M_{\oplus}$] & $\chi^2_\text{red.}$ \\
\hline
40\%     & 3.78   & $1.01 \times 10^{-2}$          & 1.41                & 3.80   & $9.41 \times 10^{-3}$          & 1.33 \\
50\%     & 3.76   & $1.08 \times 10^{-2}$          & 1.43                & 3.78   & $9.94 \times 10^{-3}$          & 1.30 \\
60\%     & 3.73   & $1.19 \times 10^{-2}$          & 1.57                & 3.76   & $1.06 \times 10^{-2}$          & 1.36 \\
70\%     & 3.68   & $1.39 \times 10^{-2}$          & 1.88                & 3.74   & $1.15 \times 10^{-2}$          & 1.51 \\
80\%     & 3.63   & $1.68 \times 10^{-2}$          & 2.49                & 3.73   & $1.28 \times 10^{-2}$          & 1.76 \\
\hline
\end{tabular}
\end{table}
In the simulations done with our model $\eta$ is always smaller, compared to those done with the Li \& Greenberg model, to counteract the steeper slope of $\kappa$ at longer wavelengths in our data set. The difference of $\eta$ between the two models grows with the amount of ice in the mixture. That is expected because the higher the amount of ice the stronger the influence of the ice data themselves on the SED. The differences are rather subtle, but not negligible. For moderate ice fractions ($<70\%$), they are less than 0.05 in terms of $\eta$, despite the big gap in $\kappa$ between our model and that from \citet{li+greenberg1998} at wavelengths greater than $200\mum$. This is due to the dominant influence of the astronomical silicate on the slope of $\kappa$ (see Fig.~\ref{fig:comparison_ice-models}) and therefore the SED.

Comparison of the observational data with the model SED produced from our optical data reveals a discrepancy at longer wavelengths that increases with the fraction of ice. This discrepancy, which is caused by the $175\mum$ feature and the resulting break in the absorption and emission efficiency, is also reflected by an increased $\chi_\text{red}^2$, as seen in Table~\ref{tab:best-fit_eta}. The SEDs modelled with data of \citet{li+greenberg1998} do not exhibit that break.

The slope $\eta$ is known to be closely related to the physics in the underlying collisional cascade from which the size distribution results. For example, a well-known relation exists between $\eta$ in a steady state and the slope of the size-dependent critical specific energy for disruption, $b$ \citep{obrien2003}:
\begin{equation}
  \eta = -\frac{21 + b}{6 + b},
\end{equation}
with $b = 0\ldots -0.4$ \citep[e.g.][]{benz+asphaug1999,wada2013}.
Assuming that a power-law size distribution is realistic, we can use the new data to rule out ice fractions above 60\%. With higher fractions of ice, the index $\eta$ of the size distribution alone can no longer be used to fit the observational data properly. While a flatter size distribution, corresponding to a lower $\eta$, would be needed to compensate the steeper fall-off in the sub-millimetre, it would also reduce the emission at shorter wavelengths, introducing a discrepancy with the IRAS and the PACS $70\mum$ data.

A more general fitting approach could still yield good agreement for ice-dominated models. An overall flatter size distribution with an additional peak of micron-sized grains would reproduce both the short- and long-wavelength observations. Such a wavy deviation from a power law might be explained readily by a collisional cascade with a lower cut-off due to radiation pressure blow-out \citep{thebault2003,krivov2006}, potentially shaped further by drag processes \citep{reidemeister2011,wyatt2011} as well as collisional spreading and damping \citep{thebault2009,pan+schlichting2012}. In that case, the additional constraints on the chemical composition would instead be transformed to constraints on the collisional physics. Thus, an accurate determination of $\eta$, or the shape of the size distribution in general, is important for understanding the material properties and variety of physical processes operating in debris disks.

With the new model, we can also give more accurate values for the modelling parameters, shown here exemplarily for $\eta$. Because evidence exists that in some cases up to 90\% of water ice is assumed to be realistic \citep[e.\,g. the Kuiper belt in the solar system,][]{vitense2010}, this would lead to significant changes in $\eta$. Besides, the colder the disk \citep[e.g. cold debris disks in][]{krivov2013}, the greater the difference, because at longer wavelengths $\alpha$ grows with decreasing temperature (see Fig.~\ref{fig:slope}). Whereas \citet{li+greenberg1998} has a constant $\alpha_\text{Li+Greenberg}=0.5$ over the entire wavelength range, our $\alpha$ is markedly higher.

Besides $\eta$, we calculated the total dust mass $m_\text{dust}$ of the disk by fitting the model to the available observations for $\lambda > 20\mum$ (Table~\ref{tab:data_HD207129}). Like $\eta$, the inferred mass depends on the ice fraction and the ice data. The mass is always a bit higher for our ice model, ranging from $1.01 \times 10^{-2} M_{\oplus}$ to $1.68 \times 10^{-2} M_{\oplus}$, and the difference to the model with the \citet{li+greenberg1998} data increases with the ice fraction.

At this point, however, we refrain from deeper analysis for that object.
%
\section{Conclusions} \label{sec:conclusions}
In this paper, we presented a new data set for the imaginary part of the refractive index $\kappa$ (describing the absorption) of crystalline water ice. The new data cover a  wavelength range of $\lambda=81\ldots625\mum$ and a temperature range of $\vartheta=10\ldots250\K$. With this, our data extend across most of the far-infrared region, including the previous gap in the data from about $200\mum$ to $235\mum$. Additionally, a complete temperature coverage is now achieved for that wavelength range. To check our results, we compared them to the previous measurements at temperatures and wavelengths, at which the latter were available, and found very good agreement with our data.

We find that for $\lambda\lesssim200\mum$ the slope of $\kappa$ stays flat and shows no change with temperature. At longer wavelengths, $\kappa$ falls off with increasing $\lambda$. In this wavelength range, the slope of $\kappa$ sensitively depends on the temperature, getting steeper at lower temperatures. This behaviour persists over the whole temperature range. The temperature dependence of $\kappa$ starts at a wavelength of $175\pm6\mum$. This can be associated with an absorption feature caused by intermolecular bending modes \citep{wozniak+dera2007} at a wavelength of around $166\mum$. This can be noted in the data compilation of \citet{warren1984}, but not in later data sets. We can now confirm these data with a direct measurement.

We approximated the data by a simple power-law based model. The model can be conveniently applied to compute absorption values for water ice at any temperature of interest.

Applying the new model to the debris disk HD\,207129 with a temperature of 55\,K results in a noticeable difference from a similar model, but with ice data from \citet{li+greenberg1998}. The derived size distribution of dust becomes flatter, while the dust mass slightly increases. The larger the assumed ice fraction in the dust composition, the greater the differences. A comparison with the observational data also gave a clue about the possible ice content in the disk, which can most probably not be higher than 70\% in the case of HD\,207129. In general, we argue that replacing the previously used ice data sets with the set presented here may result in appreciable changes of the parameters of many debris and protoplanetary disks inferred from their modelling.
%
\acknowledgements{
We thank the referee, Bernard Schmitt, for constructive remarks, which helped to improve our paper. Also we want to acknowledge Gabriele Born for  her help in the laboratory and the \textit{Deutsche Forschungsgemeinschaft} for funding our work (grants \hbox{MU1164/8-1} and \hbox{MU1164/7-2}).
}


%
\end{document}